\documentclass[
  journal=jpcafh,        
  manuscript=article,    
  layout=twocolumn,    
  articletitle=true,     
  etalmode=firstonly,    
  maxauthors=10,         
]{achemso}

\SectionNumbersOn
\usepackage[version=4]{mhchem} 
\usepackage{gensymb}
\usepackage{graphicx}
\usepackage{color}
\usepackage{placeins}

\providecommand{\doi}[1]{} 
\newcommand{\E}[1]{$\times10^{#1}$}
\newcommand{\rateUnits}{cm$^3$ molecule$^{-1}$ s$^{-1}$}
\newcommand{\terUnits}{cm$^6$ molecule$^{-2}$ s$^{-1}$}
\newcommand{\mcc}{molecules cm$^{-3}$}
\newcommand{\mz}{\textit{m}/\textit{z}}
\newcommand{\Ag}{\ce{Ag+}}
\newcommand{\OO}{\ce{O2}}
\newcommand{\AgOO}{\ce{AgO2+}}

\author{Darya Kisuryna}
\altaffiliation{These authors contributed equally.}
\affiliation[UMD IPST]
{Institute for Physical Science and Technology, University of Maryland, College Park, MD USA}

\author{Sanjana Maheshwari}
\altaffiliation{These authors contributed equally.}
\affiliation[UMD]
{Department of Chemistry and Biochemistry, University of Maryland, College Park, MD USA}

\author{Santiago Lorenzi}
\affiliation[UMD MechE]
{Department of Mechanical Engineering, University of Maryland, College Park, MD USA}

\author{Julianna Palot\'as}
\affiliation[UMD]
{Department of Chemistry and Biochemistry, University of Maryland, College Park, MD USA}
\altaffiliation{Current address: Stavropoulos Center for Complex Quantum Matter, Department of Physics and Astronomy, University of Notre Dame, Notre Dame, IN, USA}

\author{Jessica Palko}
\affiliation[UMD]
{Department of Chemistry and Biochemistry, University of Maryland, College Park, MD USA}

\author{Nathan McLane}
\affiliation[UMD IPST]
{Institute for Physical Science and Technology, University of Maryland, College Park, MD USA}

\author{Ece M. Ko\c cak}
\affiliation[UMD MechE]
{Department of Computer Science, University of Maryland, College Park, MD USA}

\author{Randall E. Pedder}
\affiliation[Ardara]
{Ardara Technologies L.P., Ardara, PA USA}

\author{Leah G. Dodson}
\email{ldodson@umd.edu}
\affiliation[UMD]
{Department of Chemistry and Biochemistry, University of Maryland, College Park, MD USA}

\title[GDIT for Kinetics]{Development of a Glow-Discharge Ion-Trap Instrument for Measuring Effective Radiative-Association Rate Coefficients}

\abbreviations{GDIT}

\begin{document}

\begin{tocentry}

\includegraphics[width=\linewidth]{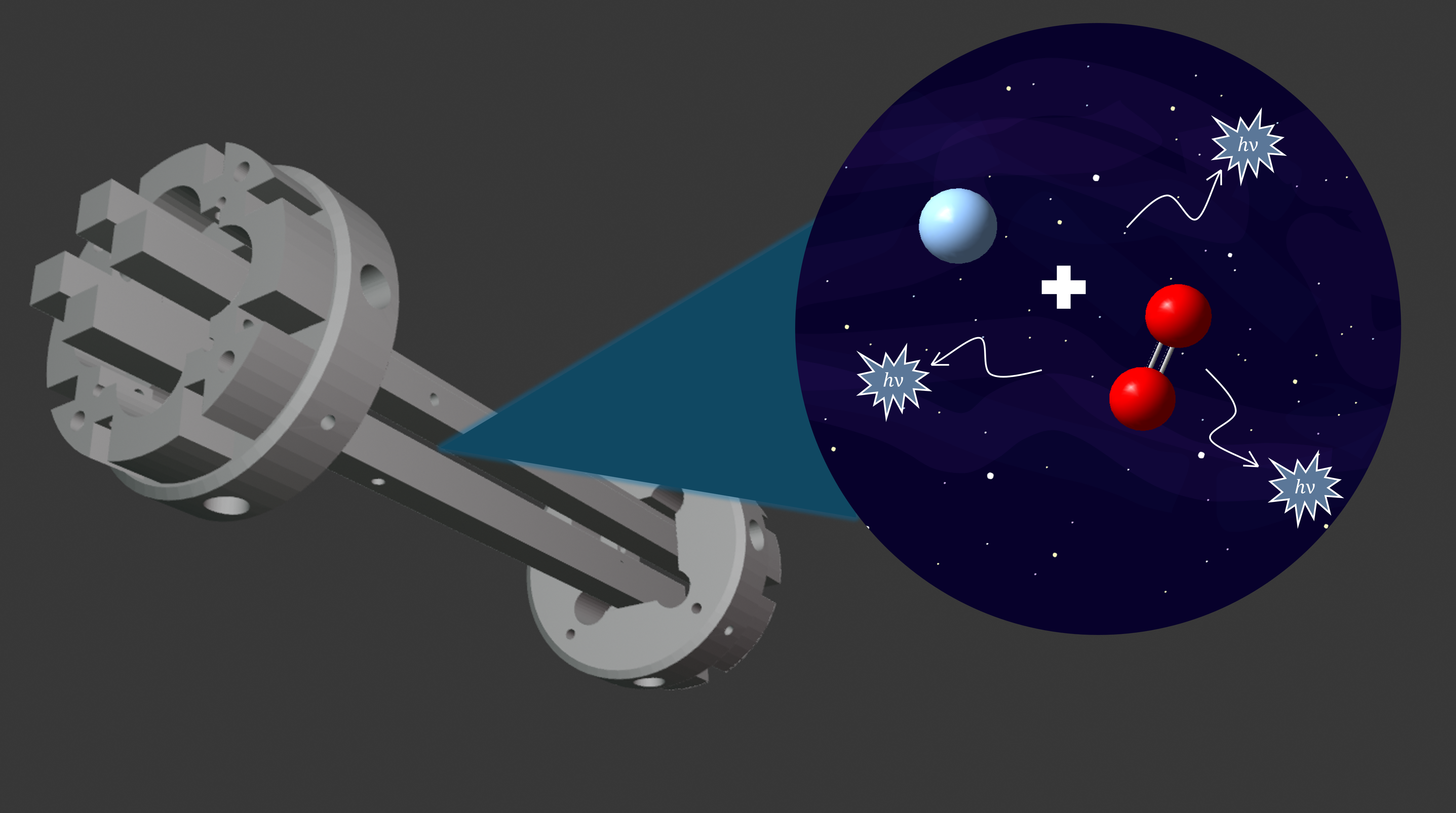}

\end{tocentry}

\begin{abstract}
  The ability to directly measure radiative-association rate coefficients for reactions between ions and neutral molecules has long challenged chemical physics laboratories, yet radiative association is one of the most important processes occurring in cold, diffuse regions of space. A reaction kinetics instrument has been developed for the investigation of ion--molecule radiative-association reactions, aimed at measuring slow, effective reaction rate coefficients for species relevant to astrophysical objects. The instrument consists of a glow-discharge ion source for production of bright and stable ion currents, a quadrupole mass filter for mass selection and detection, and a quadrupole ion trap capable of trapping reactants and products for the long times needed to measure slow kinetics. The performance and adaptability of the glow-discharge ion source has been evaluated using several configurations. To assess the feasibility of measuring reaction rate coefficients, the reaction of \ce{Ag+} and \ce{O2} was studied under pseudo-first-order conditions in the ion trap at room temperature. We present the first pressure-dependent study of this reaction and extract a lower limit of $1 \times 10^{-15}$ cm$^3$ molecule$^{-1}$ s$^{-1}$ for the \ce{Ag+ + O2} effective radiative-association rate coefficient. Measurements of effective radiative-association rate coefficients are possible for diverse atomic and molecular ions that react with neutral molecules over a range of rates in this versatile new instrument. 
\end{abstract}

\section{Introduction} \label{sec:intro}
The first molecule to appear in the universe was formed through a radiative-association reaction---an uncommon mechanism on Earth that is dominant in space.\cite{herbst_gas-phase_1989} Primordial helium hydride ion \ce{HeH+} is theorized to exist at the beginning of the universe but the molecular ion was not actually confirmed in an astrophysical object until 2019; \ce{HeH+} is thought to form via radiative association between neutral helium atoms and protons.\cite{galli_dawn_2013,gusten_astrophysical_2019}. In astrochemical models, radiative-association reactions are viewed as a precursor process, often followed by dissociative recombination, leading to the formation of neutral atoms and molecules---including but not limited to atomic \ce{H}, and metal-containing \ce{MgNC}, \ce{MgCN}, and \ce{CaCN}\cite{kawaguchi_laboratory_1993, petrie_formation_1996,dunbar_interstellar_2002,petrie_circumstellar_2004}. These reactions are important in many regions of space due to the absence of a potential barrier on the reaction coordinate since barriers are generally insurmountable in low-temperature astrophysical environments\cite{herbst_formation_1973,black_formation_1973,herbst_dynamic_1976,smith_molecular_1977,herbst_global_1991,sims_rate_1993}. From a physical chemistry perspective, radiative association is understood as a complex-forming process in which an energized collision complex may either redissociate, undergo collisional stabilization, or be stabilized through photon emission. Statistical and phase-space models, master-equation approaches, and explicit quantum dynamical scattering treatments have been developed to describe the competition between these pathways and to predict effective radiative-association rate coefficients under astrophysical and laboratory conditions.\cite{bass_ion-molecule_1981,bates_radiative_1988,Dunbar1990,klippenstein_theory_1996,Stoecklin2013,nyman2015computational}

Radiative-association reactions involving metal ions are of particular interest because atomic metal ions (\ce{Fe+}, \ce{Mg+}, \ce{Na+}, etc.) are thought to be the major carriers of positive charge in interstellar clouds and circumstellar envelopes. \cite{graedel_kinetic_1982,de_boisanger_ionization_1995,caselli_ionization_1998}. Indeed, the aforementioned metal-containing molecules in their ionic precursor form (\ce{MgNCH+}, \ce{MgCNH+}, \ce{CaCNH+}) are proposed to be formed via radiative association\cite{kawaguchi_laboratory_1993, petrie_formation_1996,dunbar_interstellar_2002, petrie_circumstellar_2004}. A deep understanding of the radiative-association reactions that form metal-containing molecules is needed to track the transfer of positive charge and other exotic chemistry across various astrophysical environments. Hence, radiative-association reactions play a dominant role in interstellar chemistry; however, direct experimental measurements in laboratory environments remain scarce.\cite{bates_radiative_1988,gerlich_experimental_1992, smith_ion_1992, gerlich2005laboratory,smith_temperature-dependence_2006,snow_ion_2008, dishoeck_astrochemistry_2014}

Radiative-association reactions are difficult to study in the laboratory due to the diffuse conditions that they generally require. Although neutral--neutral radiative-association reactions exist, most laboratory measurements involving radiative association have studied reactions between an ion and neutral pair. Barlow et al. was the first to directly measure radiative-association rate coefficients using an ion trap in 1984.\cite{barlow_radiative_1984,barlow_measurement_1986} The inevitable progress of ion-storing techniques made experimental investigations of radiative association more accessible, drawing the attention of more research groups.\cite{gerlich_experimental_1992} Experimental investigations of ion--molecule radiative association have been carried out using ion traps, flowing afterglow techniques, and beam--trap hybrids, enabling measurements of effective radiative-association rate coefficients over a wide range of temperatures and pressures\cite{gerlich_ion_1989, anicich_lifetime_1991,gerlich_experimental_1993,dunbar_radiative_1993,ryzhov_radiative_1996,luca_low_2002,smith_variable_2012,plasil_stabilization_2012}

The present work describes new instrumentation that greatly extends the availability of laboratory measurements of effective radiative-association rate coefficients of ion--molecule reactions through a modular and versatile design structure. Throughout this work, radiative-association rate coefficients refer to experimentally accessible effective rate coefficients that incorporate complex formation and competition between stabilization and dissociation. The goal of the design is to produce a bright and continuous stream of diverse ions with a well-defined energy, while also enabling measurement of very slow reaction kinetics. The glow-discharge ion-trap (GDIT) instrument enables the study of many different ions with minimal downtime, and the neutral reaction partner is only limited by volatility. Reaction timescales span from 100 ms out to at least 5 s, enabling the study of a broad range of reactivities. In Section~\ref{sec:general} we describe general considerations for the instrument design, followed by a detailed description of the instrument in Section~\ref{sec:design}. We document the overall performance of the instrument as an ion source and mass spectrometer in Section~\ref{sec:performance}, followed by a reporting of a proof-of-principle experiment studying reactions between \Ag{} ions and \OO{}. In Section~\ref{sec:reaction-kinetics}, we report the pressure-dependence of the \ce{Ag+ + O2} rate coefficient and deconstruct the total association rate into effective radiative and three-body components. Future directions and improvements are mentioned in Section~\ref{sec:conclusions}. 

\section{General Considerations} \label{sec:general}

\subsection{Choice of Ion Source} \label{sec:gen-source}

Metals of interest in our astrochemical studies, for example magnesium and aluminum, are detected in circumstellar envelopes in fairly large amounts as a part of organic neutral and ionic compounds.\cite{cernicharo_metals_1987,ziurys_more_2002,pardo_magnesium_2021,changala_laboratory_2022,cernicharo_magnesium_2023,cabezas_discovery_2023} Interestingly, these metal-bearing charged species detected in space mostly have a  charge of $+1$, whereas ions that contain magnesium or aluminum on Earth usually have a charge of $+2$ or $+3$. Our goal was to adopt an ion source capable of producing a continuous, stable stream of singly charged metal ions. Between continuous and pulsed ion sources, we give preference to the former since the goal of our experiments is to obtain reaction kinetics data, which requires a stable ion current over long experiments. The ionization technique that meets our needs is the glow discharge technique, whereas other common sources either produce multiply charged magnesium or aluminum---electrospray ionization would almost certainly involve a \ce{Mg^{2+}} salt or \ce{Al^{3+}} salt---or generally rely on pulsed operation as is the case with laser ablation/vaporization.  

The glow-discharge technique holds a strong position in high-sensitivity elemental analysis and is useful in a number of research fields: from optical emission spectroscopy to ion chemistry.\cite{harrison_glow_1986,taylor_application_1993, wilke_glow_2011} The advantages of the technique are an intense and stable ion signal, wide metal variety, mechanical simplicity, and low operational cost. 

The glow-discharge ion source is controlled by the following parameters: anode and cathode voltage, distance between the electrodes, and pressure of the inert gas. While varying those parameters, we can monitor the current between the electrodes and manipulate the formed plasma and the energetics of the resulting ions. Another advantage of the glow-discharge ion source is the physical simplicity to access the metal sample. The metal sample can be exchanged with relative ease and minimal disruption to the vacuum system, enabling the study of more than one type of metal. Finally, other atomic or even molecular ions can be produced via charge exchange if suitable precursors are added to the glow-discharge working gas.

\subsection{Quadrupole Mass Filter} \label{sec:gen-QMF}

In the interest of reducing the dimensions of the instrument as well as enabling future goals that require physical access by other instruments to the reaction cell, we elected for a design that uses one quadrupole mass filter to complete two mass-selective tasks. First, the mass filter performs mass selection of reactant ions. Although the glow-discharge ion source produces metal ions solely as monocations, other impurity ions can be produced in the source (e.g. \ce{Ar+}, \ce{O2+}, \ce{NO2+}) in the presence of minor contamination; these ions should ideally be excluded from the reaction cell. Therefore, we implemented a quadrupole mass filter prior to the reaction cell to isolate only the ion of interest. In a second step---after reactions are carried out in the reaction cell---remaining reactant ions and any formed product ions return for a second pass through the same quadrupole mass filter, enabling us to carry out mass-selective detection without adding an additional chamber or mass spectrometer. 

\subsection{Linear Quadrupole Ion Trap} \label{sec:gen-LTQ}

The ion--molecule reactions of interest are expected to have relatively slow effective reaction rate coefficients, possibly as slow as $10^{-15}$ \rateUnits{}---although $10^{-11}$ to $10^{-12}$ \rateUnits{} will likely be typical. Under reasonable reaction conditions where neutral reactants are kept at low concentrations---as low as $10^{12}$ \mcc{}---to reduce secondary reactions and clustering, the half-life of ions undergoing pseudo-first-order decay could be as long as 100 s, with typical values ranging from 100 ms to 5 s. In order to measure such slow reaction rates, the reaction conditions for the ions must be stably maintained for long times, which can be achieved in ideal scenarios in ion traps. Linear ion traps are known for their high efficiency and large storage capacities, and they can maintain stable conditions for up to several hours.\cite{church_storagering_1969,green_observation_2007, schwarz_cryogenic_2012} An additional consideration in selecting the reaction cell is the typical operating pressure, which for linear ion traps can be as low as a few mTorr.\cite{collings_combined_2001, ouyang_rectilinear_2004} Our goal is to study reactions under conditions where three-body collisions are improbable; taking this consideration, along with the expected reaction timescales, we have chosen a quadrupole linear ion trap for the reaction cell.  

\section{Design of the Instrument} \label{sec:design}

\begin{figure*}[t]
    \centering
    \includegraphics[width=\textwidth]{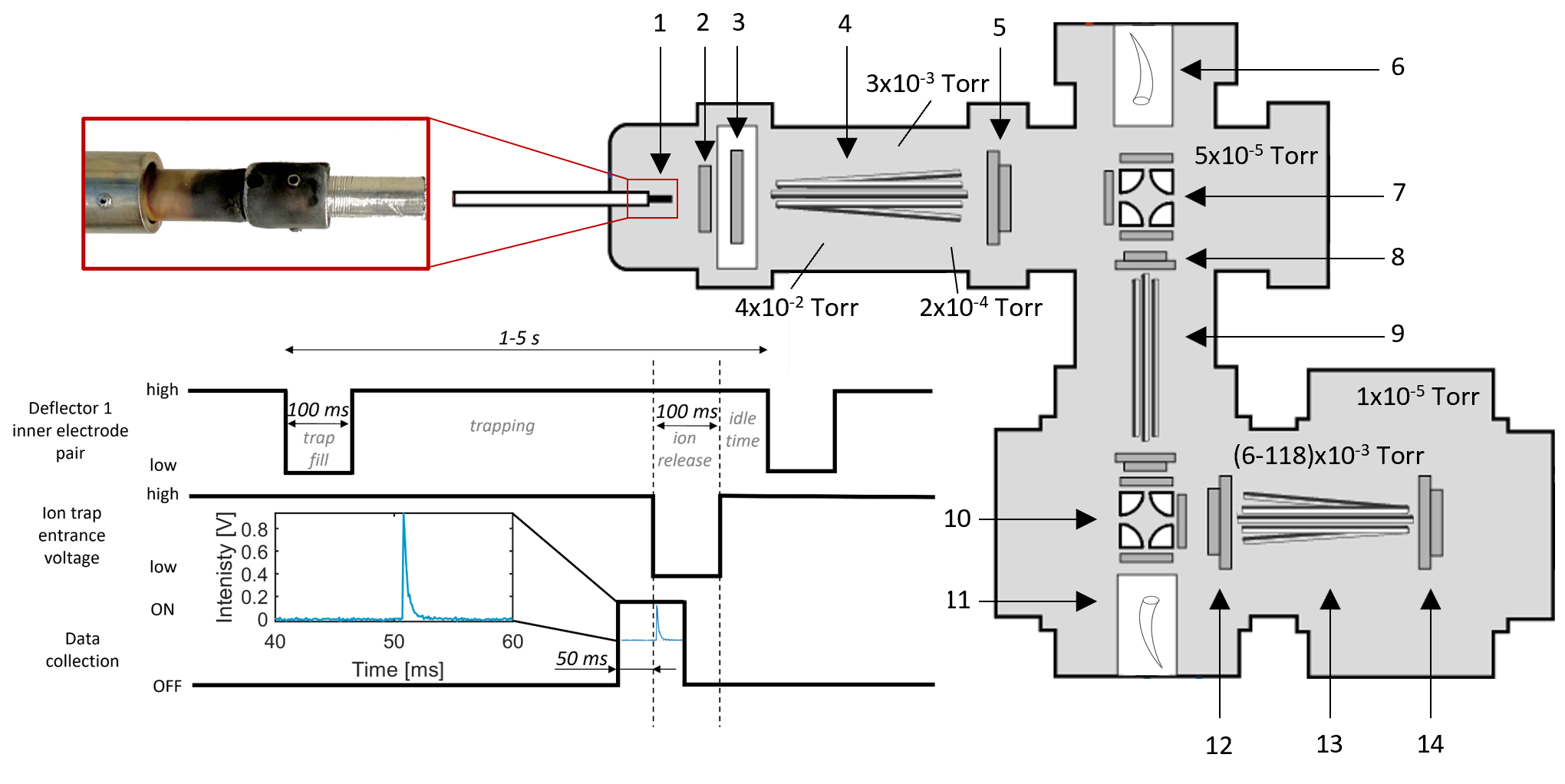}
    \caption{\textit{Right}: The GDIT instrument schematic view: (1) glow-discharge cathode, (2) exit lens serving as the anode, (3) gate-valve aperture, (4) quadrupole ion guide with asymptotic rods, (5) focusing ion lenses, (6) electron-multiplier detector 1, (7) quadrupole deflector 1, (8) focusing ion lenses, (9) quadrupole mass filter, (10) quadrupole deflector 2, (11) electron-multiplier detector 2, (12) ion-trap entrance lens, (13) ion trap with asymptotic rods, (14) ion-trap exit lens. \textit{Top left}: A photo of the glow-discharge cathode pictured with an unused silver cathode. The cylindrical metal sample is inserted into the glow-discharge wand and fixed by three set screws. \textit{Bottom left}: Timing sequence for operating the GDIT instrument. The top trace controls the first deflector electrode pair switch, which governs the trap-fill process. The second trace controls the ion-trap entrance opening time for ion extraction. The bottom trace defines the data collection window. An example ion packet obtained each duty cycle is shown in the inset window (blue trace). Not to scale.}
    \label{fig:instrument-schematic}
\end{figure*}

The schematic drawing of the glow-discharge ion-trap (GDIT) instrument is shown in Fig.~\ref{fig:instrument-schematic} demonstrating the general layout and individual components. GDIT consists of a glow-discharge ion source for the stable ion generation necessary for performing kinetics experiments, a quadrupole mass filter for mass selection of both the reactant and the product ions, and a linear ion trap for containing ions during ion--molecule reactions and measuring reaction kinetics on the relevant reaction timescales.

\subsection{Glow-Discharge Ion Source and Quadrupole Mass Selection} \label{sec:det-source}

\begin{figure}[t!]
    \centering
    \includegraphics[width=3.33in]{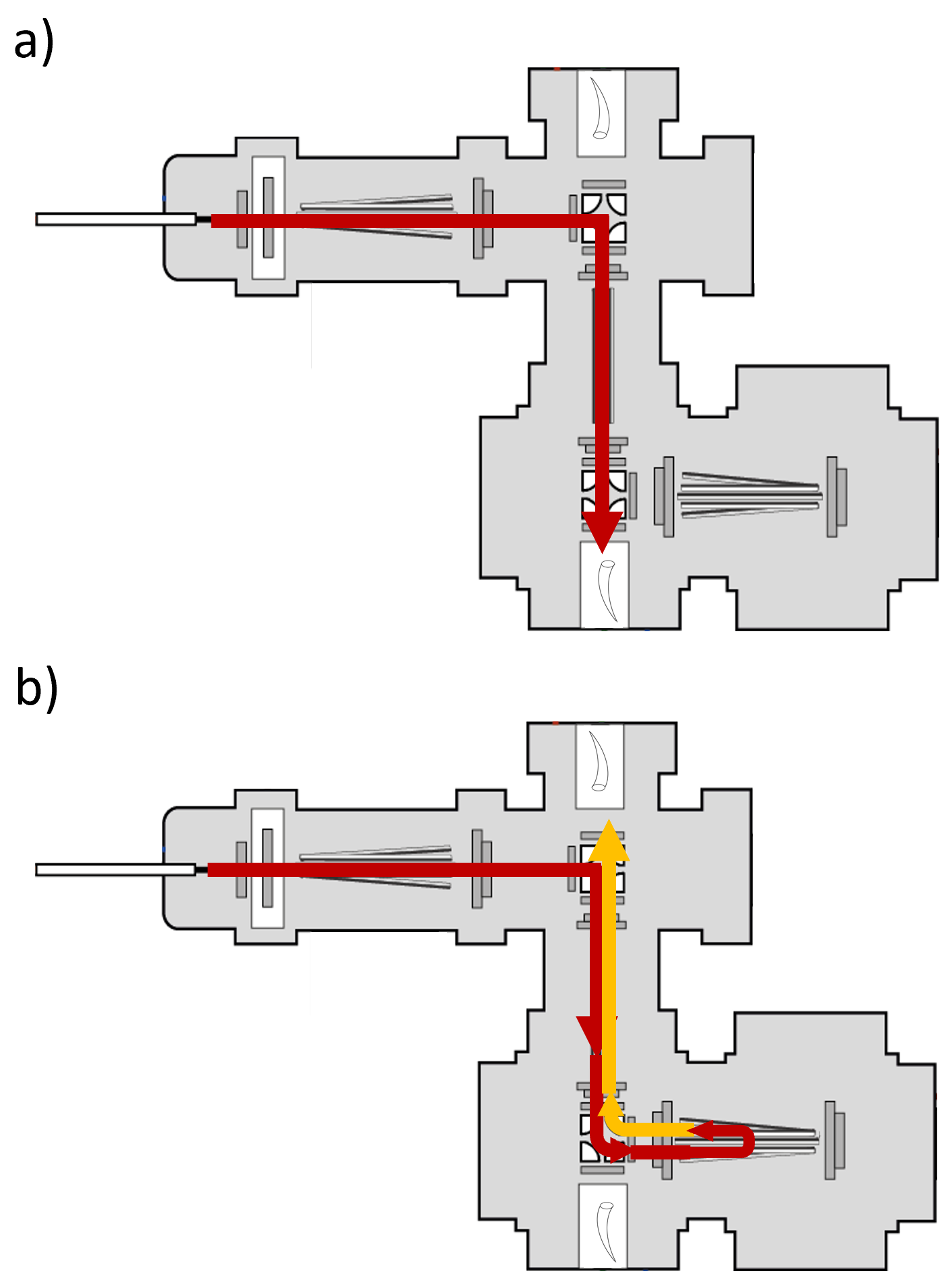}
    \caption{Ion trajectories for \textit{a}) mass spectrum collection; \textit{b}) kinetics experiments. In both configurations after generation in the glow-discharge chamber the reactant ions travel following the red trace through the ion guide, bend 90\degree{} via deflector 1 and are \textit{a}) mass resolved or \textit{b}) mass selected in the quadrupole mass filter. While in \textit{a}) ions culminate at electron-multiplier detector 2, in \textit{b}) ions bend 90\degree{} by deflector 2 towards the ion trap. In \textit{b}), after a pre-selected time interval both reactant and product ions exit the trap and travel following the yellow trace through deflector 2 towards the quadrupole mass filter to be mass resolved and detected by electron-multiplier detector 1.}
    \label{fig:ion-trajectories}
\end{figure}

The glow-discharge ion source was custom designed by Ardara Technologies specifically for this system. The sample probe is constructed from a $1/2''$ outer-diameter polished stainless steel tube with a coaxial center electrode. The coaxial center electrode is electrically shielded from the outer tube along its length by an aluminum-oxide ceramic tube to prevent discharge with the probe itself. At the tip of the center electrode a $1/4''$-diameter, $1''$-long cylindrical sample rod is attached, see inset in top left of Fig.~\ref{fig:instrument-schematic}.

The volume around the final portion of the sample probe is terminated by an aluminum-oxide ceramic washer. Immediately after this conductance-limiting insulating washer is a stack of stainless steel spacers that form the anode electrode. The length of this volume can be varied by changing the number of these stainless steel spacers to optimize the emissivity of the glow discharge. The last of these stainless-steel electrodes serves as the conductance-limiting aperture (Fig.~\ref{fig:instrument-schematic} (2)) where ions are sampled from the glow-discharge ion source into the next vacuum region. The sample probe is inserted into the glow-discharge chamber separated from the rest of the assembly by the gate valve. The gate valve (3) configuration allows easy replacement of the sample without venting the whole system.

The plasma gas---ultra-high purity (UHP) argon (Airgas, 99.999\%)---is introduced into the glow-discharge chamber, ultimately flowing into the small glow discharge region surrounding the sample probe pin. In this way, there is coaxial flow of the plasma gas, with its only exit being the conductance-limiting exit aperture of the anode stack. This gas-flow configuration constantly flushes the finite plasma volume, allowing for fast equilibration of the plasma after a new sample pin is inserted. Gas pressure in the glow-discharge cell is maintained by a needle valve and set to about 1 Torr as measured on a gas-type-calibrated convection gauge (Instrutech CVG101 Worker Bee) affixed to a flange on the side of the glow-discharge volume. The applied high voltage is provided by a high-voltage power supply (Gamma RR3-25R) that can be tuned continuously up to $\pm3.0$ kV. The high voltage causes the breakdown of the inert gas medium resulting in the formation of \ce{Ar+}, metastable excited atoms, and electrons. Then Ar cations bombard the high-voltage cathode (1) causing surface sputtering and metal-atom ejection.\cite{taylor_application_1993} The distance between the electrodes is variable ($0$ to $0.8''$), as is the Ar gas pressure; both parameters play a crucial role in the establishment of the plasma current and the stability of the ion signal. 

The ions leave the glow-discharge cell through the exit lens (Fig.~\ref{fig:instrument-schematic} component (2)), followed by the gate-valve aperture (3) that enables isolation and removal of the ion source for cleaning or cathode exchange. The glow-discharge anode aperture size can be varied by installing the required ion lens and was found to produce optimal signal at 1-mm diameter. Ions then pass through the RF-only rectilinear quadrupole ion guide with asymptotic rods (4). This RF-only ion guide with 5-mm inscribed diameter traverses three stages of differential pumping, maintained by an Edwards nEXT cartridge pump with 20, 200, and 200 L/s pumping speeds for the three pumped regions. Nominal pressures in these three regions are ca. 40 mTorr, 33 mTorr, and 0.2 mTorr respectively, exiting to a quadrupole deflector energy filter. 

The ions may then be deflected by a quadrupole deflector energy filter (7) either to the electron-multiplier detector (6) or after being focused by a set of ion optics (8) into a quadrupole mass filter (9) where mass selection is achieved during this first pass. The mass spectrum of pristine ions produced in the glow-discharge source can be collected by allowing ions to pass straight through a second quadrupole deflector (10) to an electron-multiplier detector (11) (Fig.~\ref{fig:ion-trajectories}a), or the ions can be deflected 90\degree{} toward the reaction trap (13) for subsequent reaction and analysis (Fig.~\ref{fig:ion-trajectories}b). The mass axis of the mass spectrometer is primarily calibrated using the isotopes of atomic metal ions \ce{Ni+}, \ce{Cu+}, and \ce{Ag+} produced from individual pure sputtering targets, but the calibration is extended to higher masses by adding a small amount of xenon to the glow-discharge working gas and measuring the \ce{Xe+} isotope pattern.

The six-way cross chamber that houses the first deflector (7) is pumped by a 400 L/s turbomolecular pump (Edwards). Typical working pressure is 5\E{-5} Torr in the six-way cross chamber that houses the first quadrupole deflector (7). All pressures mentioned here were measured using hot cathode ionization gauges (Instrutech IGM401 Hornet).

\subsection{Reaction Trap} \label{sec:det-trap}

Chemical kinetics experiments studying reactions between ions and neutral reactant molecules are carried out in the reaction trap. Mass-selected ions that are directed 90\degree{} by the second deflector (10) are then guided to a rectilinear quadrupole ion trap (13) where trapped ions undergo chemical reactions with a reactant gas. Product ions created from the reaction, as well as the remaining reactant ions in the trap, are then sent back through the instrument, following the path in yellow in Fig.~\ref{fig:ion-trajectories}b, where they are deflected 90\degree{} (10), mass analyzed using the quadrupole mass filter (9), and are finally passed straight through (7) where they are detected as a function of \mz{} by an electron-multiplier detector (6). 

A $5.428''$-long rectilinear quadrupole ion trap, with 8-mm inscribed diameter, is contained within a conductance-limiting housing with a $1/8''$ Teflon gas inlet, and a $1/4''$ Teflon pressure-measuring port, both plumbed to a vacuum flange. A key feature of this rectilinear ion trap is the use of four additional electrodes located at the asymptotes of the rectilinear ion trap, situated such that the inside corners have an inscribed diameter that is larger at the entrance than at the opposite end of the ion trap. Applying a ``repelling'' DC voltage simultaneously to these four asymptotic field shaping electrodes effectively creates a sloped centerline potential, causing collisionally damped ions to accumulate in the portion of the ion guide at the lowest centerline potential, which is just inside the entrance lens.\cite{wilcox2002improved,yan2017surface}  

The ion trap is mainly filled with UHP Helium (He) buffer gas (Airgas, 99.999\%) that flows continuously into the ion trap from a mass flow controller (Alicat, 0--100 sccm). To carry out chemical reactions, a small amount of the buffer gas is replaced by a neutral gaseous reactant molecule by preparing a dilute stock cylinder, the contents of which are delivered to the trap using a second mass flow controller (Alicat, 0--100 sccm). The ion-trap chamber is pumped by a 400 L/s turbomolecular pump (Edwards) yielding a base pressure of 3\E{-8} Torr in the chamber, which rises to 1\E{-5} Torr during experiments. Pressure in the ion trap is measured directly by a convection gauge (Instrutech Worker Bee) connected by a $1/4''$-diameter tube to the ion-trap housing. The uncertainty in the reaction pressure is dominated by the $\pm10\%$ measurement accuracy quoted by the manufacturer, which is the main source of systematic uncertainty for our kinetics measurements. Steady-state conditions are established in the ion trap with pressures maintained in the 6 to 118 mTorr pressure range---of which less than 1\% is the reactant gas and the remainder is helium. 

\subsection{Overall Operation Procedure} \label{sec:det-steps}

A key consideration in designing this instrument was the implementation of a timing scheme (Fig.~\ref{fig:instrument-schematic} \textit{bottom left}) that enables ion detection (both reactants and products) as a function of reaction time. Broadly speaking, the trapping time of the ion trap, defined throughout as $t$, is approximately equal to the reaction time. Ions enter the ion trap where they encounter a neutral reaction partner present at a steady-state concentration that is many orders of magnitude larger than the ion density (pseudo-first-order conditions). Reaction progression is monitored by varying the amount of time ions spend in the trap before they are extracted---observing both the decay of reactant ions and the growth/subsequent decay of product ions. These measurements require a sequence of steps to transport ions through different regions; therefore, several ion optical elements in the instrument have their voltages switched from ``trap fill'' to “trapping” to “ion release” modes of the instrument, which are precisely timed by a digital delay generator (Quantum Composers 9214+) controlled by a home-built data acquisition program (LabVIEW).  

In practice, the metal ions of interest are first produced from a pure sample composed of that metal, although in principle any conductive sample will work. The glow-discharge cathode (1) is usually supplied with $-1$ to $-2$ kV depending on the metal. Cations produced from the cathode after argon bombardment pass through the exit lens (2), the focusing gate valve aperture (3), and enter the ion guide (4) where they are transferred through several stages of differential pumping. The ion guide is biased at $+10$ V in a relatively high-pressure region of the instrument, which defines the initial reference potential energy of the cations. 

During the ``trap fill'' phase of the experiment, the ions travel through a set of focusing lenses (5) before they are deflected 90\degree{} to the right by the first deflector (7) and pass through a focusing lens (8) to the quadrupole mass filter (9). At this stage of the experiment, the mass filter is optimized so that only the reactant ion of interest passes through. At the next ion deflector (10), cations are bent 90\degree{} to the left to the ion trap entrance lens (12) held at a positive voltage low enough for ions to overcome a  small potential hill---essentially acting as a speed bump. The presence of a speed bump prevents the ions from spontaneously escaping the trap after thermalization begins to occur.

After waiting for ions to accumulate in the ion trap for 100 ms, the instrument is switched to the ``trapping'' mode. The voltages applied to the first ion deflector (7) are switched so that ions from the source travel straight through (i.e., are no longer directed toward the ion trap) and are removed from the system---promptly ending the ``trap fill'' stage of the experiment. Ions already present in the reaction trap rapidly thermalize via collisions with the buffer gas and become physically trapped while keeping the entrance lens (12) positive. In the trap, reactant ions meet neutral molecules where chemical reactions can occur. Any charged products that form remain in the trap due to its wide acceptance range, which allows ions of a variety of masses to be held in the trap. The amount of time the ions remain in the trap ranges from $t=50$ ms to 5 s, as controlled by the digital delay generator and data acquisition program. The trapping time is usually limited by the repetition rate of the timing sequence, which can be changed to accommodate longer or shorter reaction times for slower or faster reactions. 

After waiting for the pre-selected trapping time, the instrument is switched to the ``ion release'' mode of operation. Ions are released from the reaction trap by switching the voltage on the entrance lens (12) to low voltage. The ions then follow their original path but in reverse, first being bent 90\degree{} to the right by the second deflector (10) then traveling through the quadrupole mass filter (9) before passing straight through the first deflector (7) where they are ultimately detected by the electron multiplier (6). After the ``ion release'' stage of the experiment is complete, the instrument enters an idle time until the end of the duty cycle with voltages reset the same as the ``trapping'' phase, before beginning the sequence again with a new ``trap fill.'' Ions arrive at the detector as a discrete packet about 1 ms wide (see inset of Fig.~\ref{fig:instrument-schematic}). Typically twenty ion packets are recorded for each independent variable setting and ion intensities are reported as the average of the peak of each ion packet, with standard deviations reflecting the random fluctuations in ion-packet intensities. 

There are two methods of operation implemented for the mass filter during the ``ion release'' mode of the instrument. The mass filter may either be scanned to obtain the ion signal at a single trapping time as a function of \mz{} (spectrometry mode) or it can have its parameters fixed so that a selected \mz{} is monitored as a function of trapping time (kinetics mode). Three-dimensional data sets can be assembled with ion signal as a function of both \mz{} and trapping time, enabling a full evaluation of the ongoing chemical reaction. In practice we operate in kinetics mode monitoring each species independently as a function of time, using spectrometry mode only to identify and monitor the products. 

\section{Performance of the Instrument} \label{sec:performance}

Here we report our first results obtained using the GDIT instrument. The ion source is first evaluated using the quadrupole mass filter as a conventional mass spectrometer. Then a kinetics experiment verifies the use of reaction trap for time-dependent measurements that include direct detection of product ions. The performance of the instrument is shown to be well-matched for studies of effective radiative-association reactions. 
\newline

\subsection{Ion Production and Detection} \label{sec:perf-production}

\begin{figure}[tbp]
    \centering
    \includegraphics[width=3.33in]{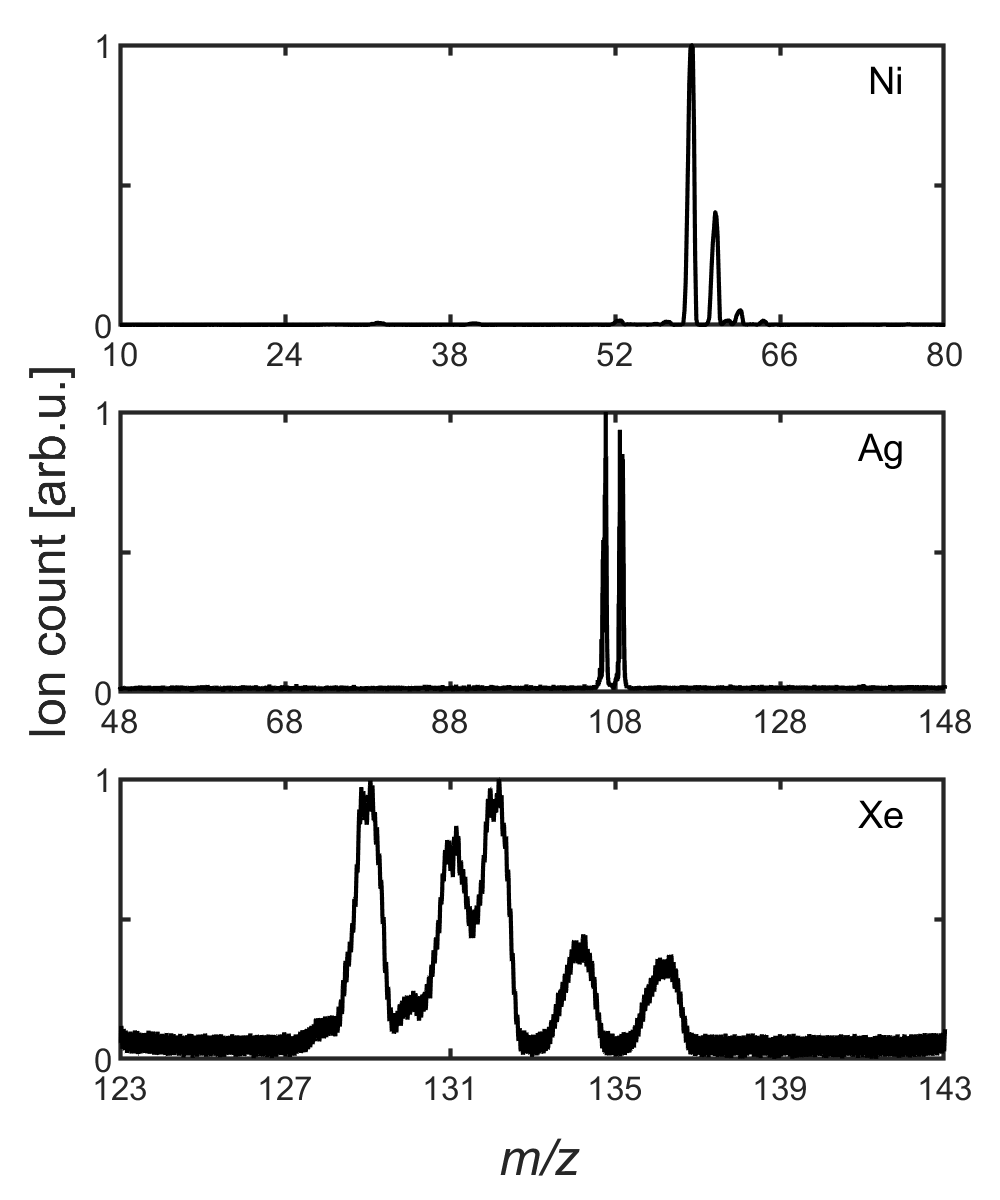}
    \caption{Mass spectra of produced ions using (\textit{top}) nickel, (\textit{middle}) silver, and (\textit{bottom}) xenon precursors.}
    \label{fig:metal-MS}
\end{figure}

To evaluate the adaptability and stability of the glow-discharge ion source, we tested several cathode materials using the simple mass-spectrometry configuration shown in Fig.~\ref{fig:ion-trajectories}a. Fig.~\ref{fig:metal-MS} shows the mass spectra measured using metal cathodes of different compositions: nickel and silver, and the spectrum obtained with xenon present in the glow-discharge working gas. Each spectrum has been normalized to the most-intense peak and can be seen to be free from impurities with the correct isotope pattern for the natural-abundance samples used in each case. The top spectrum of Fig.~\ref{fig:metal-MS} demonstrates that a cathode made from nickel easily sputters with good signal-to-noise for the four isotopes of nickel, \mz{} = 58, 60, 62, 64. The middle panel shows two resolved isotopes of silver, \mz{} = 107 and 109 produced with a silver cathode. The bottom panel shows the high-mass part of the spectrum revealing the isotopes of xenon, \mz{} = 128, 129, 130, 131, 132, 134, and 136, produced by sputtering a nickel cathode in the presence of a small amount of xenon gas. Some of the peaks in these spectra appear partially split, which might be caused by a periodic motion of ions through quadrupole mass filter, excessive energy of ions, and/or a finite exit aperture size after the mass filter.\cite{pedder_practical_nodate-1} We see no evidence of multiply charged ions in our system. When operating in this configuration, the resolution of the quadrupole mass filter can be optimized between $m/\Delta m=$ 60 to 300, which is sufficient resolution for these experiments. 

Important for kinetics experiments is long-term stability of the ion signal that persists for a complete set of experiments. We evaluated the stability over 6 hours. After an induction period of about 3 hours, the ion signal stabilizes and maintains a constant intensity with random fluctuations about 10\% over the course of a single kinetics experiment. We compensate for this minor variability by randomizing the order in which each trapping time is measured during kinetics. 

\subsection{The Ion-Trap Reactor: Reactant and Product Identification} \label{sec:reaction-products}

The primary goal of the GDIT instrument is to study the kinetics of radiative-association reactions, which firstly requires a means to distinguish complex-forming and direct (non-association) product channels via reaction product identification. An advantage of mass spectrometry is that any charged reactant and/or product within the appropriate mass range can be detected with high sensitivity, enabling mechanistic information to be obtained. As a benchmark reaction we chose the reaction between silver cation \ce{Ag+} and molecular oxygen \ce{O2} due to proceeding dominantly through an associative reaction channel (reaction~\eqref{rxn:overall}) under the conditions studied and the availability of reference data for the apparent binary reaction and rate coefficient $k^\ast$ reported by Koyanagi et al\cite{koyanagi_gas-phase_2002}. 
\begin{equation}
\ce{Ag+ + O2 ->[$k^\ast$] AgO2+}
\label{rxn:overall}
\end{equation}

The glow-discharge ion source enables collection of a clean and well-resolved spectrum for both sliver isotopes (Fig.~\ref{fig:metal-MS}). The energies of the glow discharge are well-suited to selectively produce the ground state of atomic silver cations. Experiments involving the ionization of various metals and molecular ions (e.g. nickel, magnesium, zinc, oxygen, water, etc.) indicate that the maximum achievable energies under these conditions are approximately 13 eV.\cite{kramida_nist_1999} Achieving the second ionization or excited electronic state silver would require higher discharge voltages than those used in this work. Therefore, we assume the Ag reactant ions are in their $^1$S ground state in the present work. 

\begin{figure}[h]
    \centering
    \includegraphics[width=3.33in]{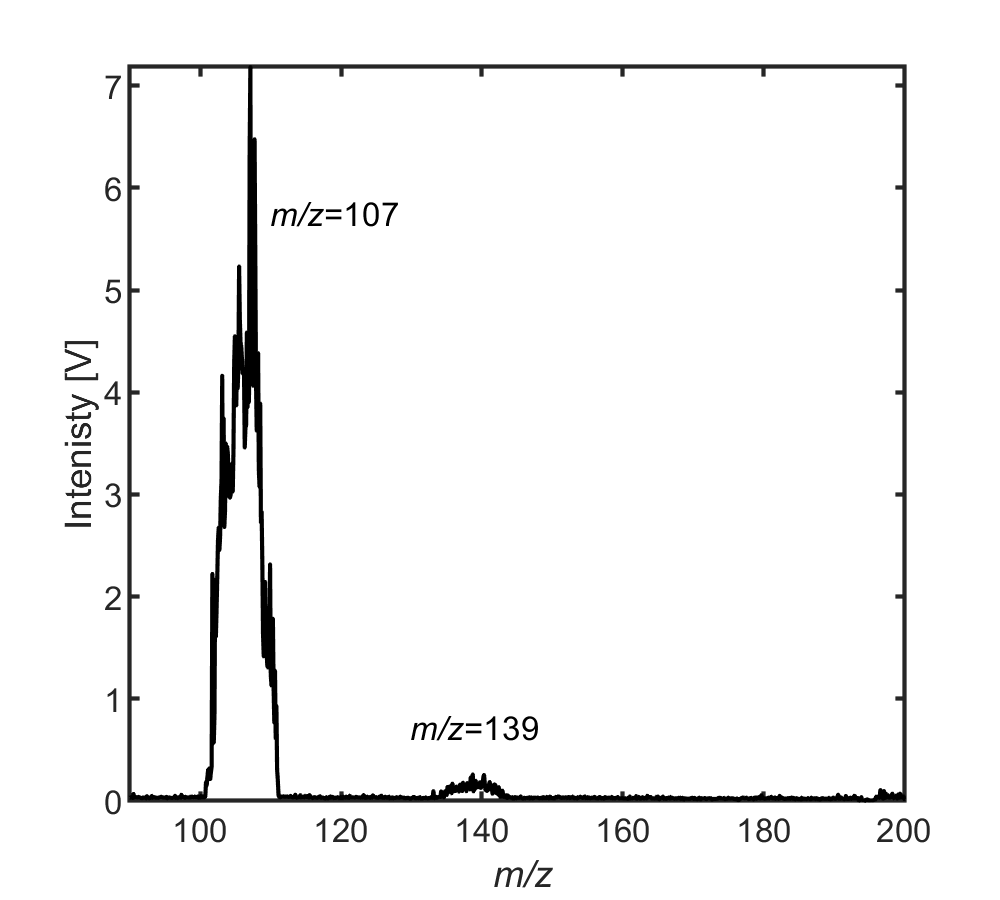}
    \caption{Example mass spectrum resulting from reaction between \Ag{} and \OO{} for identification of reactant and product ions. The mass spectrum is collected for a 2\% mixture of \OO{} diluted in He at room temperature and  17 mTorr in the trap with a trapping time of 0.5 seconds. The peak centered at \mz{} = 107 is assigned to the \Ag{} reactant and \mz{} = 139 corresponds to the \AgOO{} product.}
    \label{fig:AG_O_MS}
\end{figure}

Fig.~\ref{fig:AG_O_MS} shows the mass spectrum of the mixture of reactants and products produced from the \ce{Ag+ + O2} reaction occurring in the GDIT reaction trap at room temperature. To generate this spectrum, \ce{^{107}Ag} ions produced in the glow-discharge source were mass selected by the quadrupole mass filter and then guided into the ion trap that was filled with a steady-state flow of \ce{O2} (2\%) and \ce{He} gas. Reactants ions and any formed products were held in the trap for 0.5 seconds at constant pressure (17 mTorr) after which they were released from the ion trap and were detected as a function of \mz{}. Two distinct peaks are observed in the mass spectrum, centered at \mz{} = 107 and \mz{} = 139. The peak at \mz{} = 107 has already been identified during calibration experiments (Section \ref{sec:perf-production}) as \ce{Ag+} from mass spectra obtained in the absence of \OO{}. The mass of \mz{} = 139 corresponds to the chemical formula \ce{[AgO2]+}. An analogous mass spectrum was collected using the mass filter to selectively fill the ion trap with ions of the Ag-109 isotope, which shifted the center of the \ce{AgO2+} product peak to \mz{} 141, verifying the presence of silver in the product compound. We observed no other products in these \ce{Ag+ + O2} experiments up to the mass limit of \mz{} = 200, even at trapping times out to 5 s, and specifically note the absence of the direct bimolecular product \ce{AgO+}, in good agreement with Koyanagi et al.\cite{koyanagi_gas-phase_2002} Increasing the steady-state \OO{} partial pressure resulted in heavier-mass species (\ce{[Ag(O2)_n]^+}), which are assumed to be from higher-order chemistry.

The ability to directly observe and mitigate the contribution of secondary products and higher-order effects is an important outcome of GDIT's product-detection capabilities. The spectrum presented in Fig.~\ref{fig:AG_O_MS} was obtained under conditions specifically designed to increase sensitivity of higher-mass products. For this spectrum, mass filter settings were used that result in lower mass resolution ($m/\Delta m = 10$), which was necessary to maximize detection of reaction products. Since the chemical system under investigation for this work is relatively simple, the mass resolution was sufficient to identify each chemical species. Future improvements will be identified in Section~\ref{sec:conclusions}.

\subsection{The Ion-Trap Reactor: Reaction Kinetics} \label{sec:reaction-kinetics}

\begin{figure}[h]
    \centering
    \includegraphics[width=3.33in]{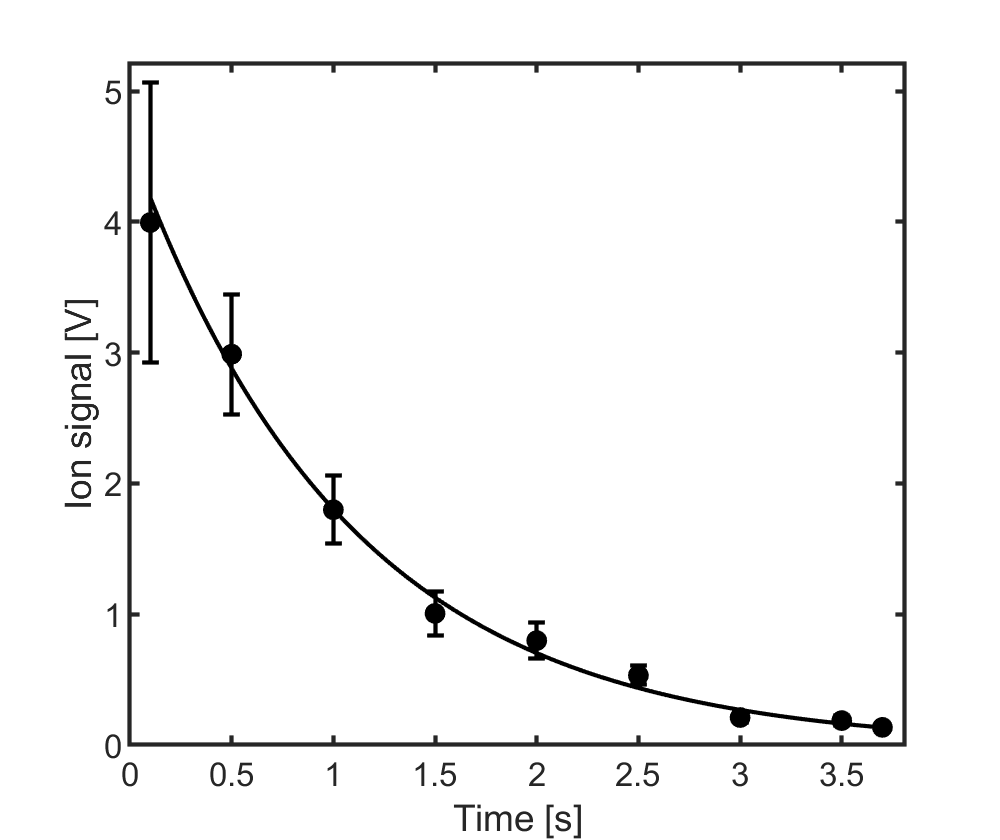}
    \caption{Kinetic trace of \ce{Ag+} depicting reactant decay in the presence of \OO. Ion packet peak voltages are accumulated and averaged over 20 samples with error bars showing uncertainty due to random fluctuations. Measurements obtained for a 0.35\% mixture of \OO{} diluted in He to a trap pressure of 18 mTorr at room temperature. The solid line is the result of an exponential fit to the data yielding $k' = (0.93\pm0.13)$ s$^{-1}$.}  
    \label{fig:Ag_decay}
\end{figure}

To experimentally measure effective radiative-association reaction rate coefficients, GDIT is designed to monitor charged reactants and products as a function of trapping time. In practice, effective rate coefficient information is largely obtained from fits of the exponentially decaying reactant ions under pseudo-first-order conditions. In studying the \ce{Ag+ + O2} reaction system, we monitored the decay of \Ag{} ions as a function of time, shown in Fig.~\ref{fig:Ag_decay} for one set of conditions.

Experiments were first conducted to verify that Ag ions remain in the trap for sufficiently long times in the absence of co-reactants. To quantify the non-reactive trap loss, we first monitored the decay of silver ions at a fixed trap pressure with only helium present. This measurement provides an upper limit for the ion trap loss rate, $k_\text{L}<0.3$ s$^{-1}$ at that pressure. In the presence of \OO{}, our experimentally observed reaction decay traces characterized by the rate $k'$ include both this first-order loss ($k_\text{L}$) as well as reactive loss of the Ag ions ($k^\ast$). Hence, the observed exponential decay of the reactant ion signal is caused by depletion of \ce{Ag} ions upon reaction with \OO{} and non-reactive loss in the ion trap. In pseudo-first-order kinetics experiments, there is sufficient \OO{} present to cause all \ce{Ag} ions to be consumed by $t = 3.5$ s under these conditions. We therefore assumed the time-dependent \Ag{} signal ($[\Ag](t)$) could be fit as a function of trapping time $t$ by the following simple model:
\begin{equation}
    [\ce{Ag+}](t) = [\ce{Ag+}]_0 e^{-k't}
    \label{eqn:exponential}
\end{equation}
where $k'$ is the pseudo-first-order rate coefficient $k'=k^\ast[\ce{O2}]+k_\text{L}$, $k^\ast$ is the apparent binary rate coefficient described by reaction~\eqref{rxn:overall}, $k_L$ is the trap loss rate defined above, and $[\Ag]_0$ is the initial \ce{Ag} ion signal at $t= 0$ s. Since there is a small 100 ms delay between the arrival of the first \Ag{} ion in the ion trap during the ``trap fill'' part of the sequence and the start of the ``trapping'' time clock, this creates a slight offset in the true time axis, which is negligible when considering the overall timescale for the slow reactions GDIT is designed to study. The relative decay rate observed starting from $t=0$ s still accurately reflects the \ce{Ag+ + O2} reaction progression and can be used to extract $k'$. For the example shown in Fig.~\ref{fig:Ag_decay}, the exponential decay rate was found to be $k' = (0.93\pm0.13)$ s$^{-1}$, where the error reflects uncertainty in the weighted fit. The quantitative measurement of such reasonably long reactant half lives reinforces the need for a measurement tool (the ion trap) that supports slow effective kinetics.

\begin{figure}[h!]
    \centering
    \includegraphics[width=3.33in]{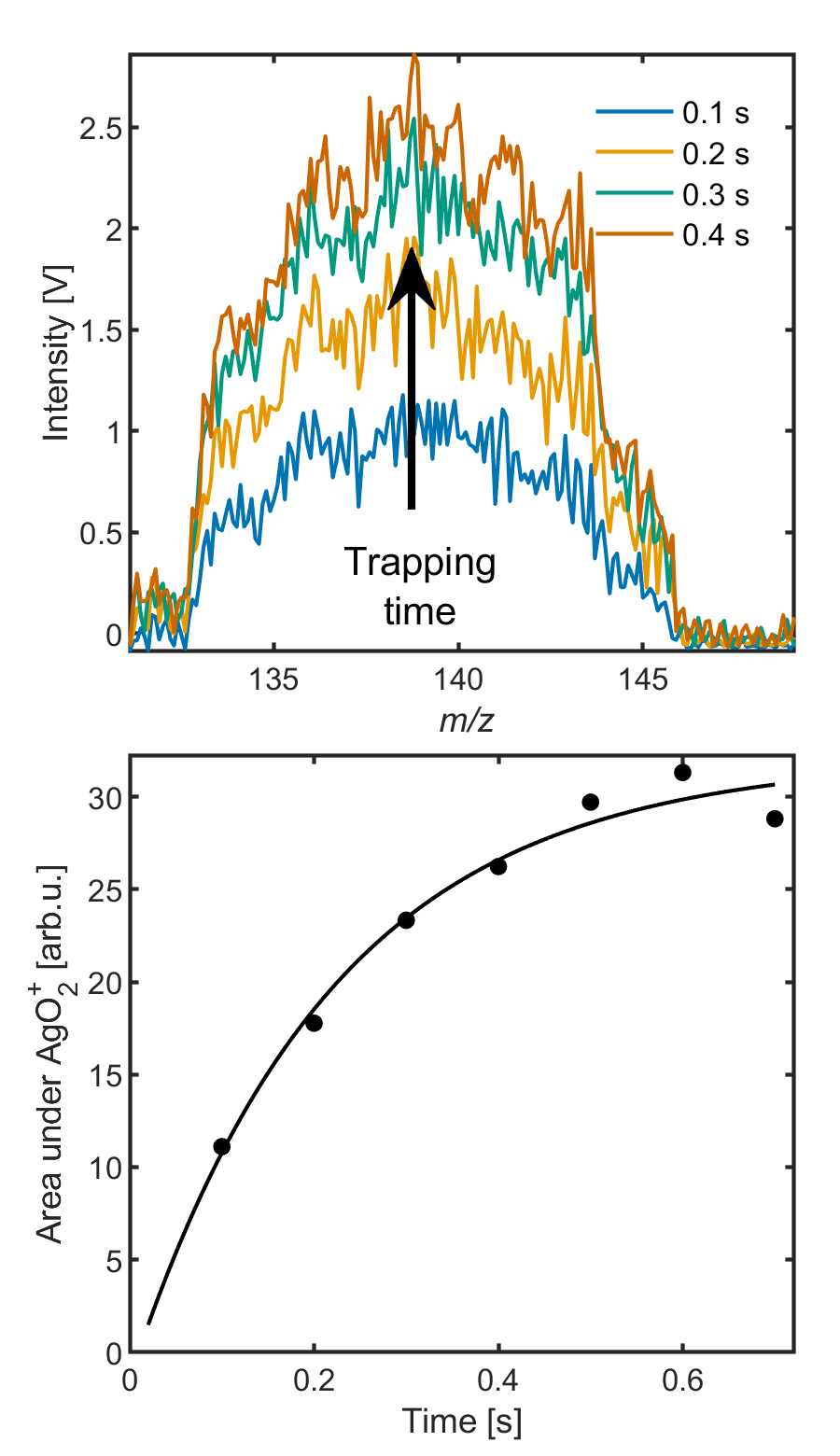}
    \caption{Kinetic trace of \AgOO{} showing product growth from \ce{Ag+ + O2}.  Data points are derived from integration of the product-peak areas in the mass spectra. One mass spectrum scan was collected for each trapping time. Measurements obtained for a 0.35\% mixture of \OO{} diluted in He to a trap pressure of 18 mTorr at room temperature.}  
    \label{fig:AgOO_growth}
\end{figure}

Product growth is also observed as a function of time in this system by monitoring the appearance of \AgOO{}. The top panel of  Fig.~\ref{fig:AgOO_growth} shows the evolution of the product centered at \mz{} 139 with an increase in trapping time. The area under product peaks was then integrated and plotted against the trapping time (bottom panel of Fig. ~\ref{fig:AgOO_growth}). The time-dependent increase in \AgOO{} peak area is fit to an exponential function to obtain a formation rate of 0.22 s$^{-1}$, which is about a factor of 3 slower than the decay rate of \Ag{} under the same conditions. The slower-than-expected growth rate is likely due to secondary reactions that consume \AgOO{}, which are already anticipated based on the prominence of \ce{[Ag(O2)_n]^+} at higher \OO{} concentrations. Taken together, the \Ag{} decay, \AgOO{} growth, and relatively clean product spectrum unambiguously demonstrate that \Ag{} reacts with \OO{} to  form \AgOO{} via a complex-forming process.  

We obtained the pseudo-first-order rate coefficient $k'=(0.93\pm0.13$)  s$^{-1}$ for \Ag{} decay (Fig. \ref{fig:Ag_decay}) at $[\OO{}] = 1.89 $\E{12} \mcc{}. This rate coefficient was then corrected for the ion trap loss rate $k_\text{L}$ and divided by the number density of [\OO{}], $k^*=(k'-k_L)/[\OO{}]$, yielding the apparent binary rate coefficient of $k^\ast=(3.2\pm0.7)$\E{-13} \rateUnits{} at 18 mTorr total pressure, where the uncertainty in $k^\ast$ is from the weighted exponential fit of the \Ag{} ion intensity as a function of time, using the random intensity fluctuations as the weighting function. 

To elucidate the reaction mechanism for \ce{Ag+ + O2}, we conducted a study of the pressure dependence of the rate coefficient. The reaction begins with the reversible formation of an activated complex:
\begin{equation*}
    \ce{Ag+ + O2 <=>[$k_f$][$k_b$] [AgO2+]^{\ast}}
\end{equation*}
The fate of the activated complex has two forward reaction pathways: stabilization to the \AgOO{} product via collisions with a third body M \eqref{rxn:Ter}---in this work assumed to be only the He bath gas---or stabilization through a radiative process \eqref{rxn:Rad}. 
\begin{equation}
    \ce{[AgO2+]^{\ast} + M ->[$k_3$] AgO2+ + M}
    \label{rxn:Ter}
\end{equation}
\begin{equation}
    \ce{[AgO2+]^{\ast} + O2 ->[$k_\text{RA}$] AgO2+ + h\nu}
    \label{rxn:Rad}
\end{equation}

The contribution of radiative association only becomes important at sufficiently low pressures so that the timescale of \eqref{rxn:Rad} is competitive with termolecular stabilization via \eqref{rxn:Ter}. In the limit of zero pressure, the only association mechanism would be \eqref{rxn:Rad}, whereas three-body association quickly dominates when collision partners are available. At intermediate pressures, the effective rate coefficient for radiative association ($k_r^\text{eff}$) can be extracted along with the effective three-body rate coefficient ($k_3^\text{eff}$) from the pressure dependence of $k^\ast$. \cite{gerlich_experimental_1992}
\begin{equation}
    \frac{\mathrm{d}[\Ag]}{\mathrm{d}t} = -k^\ast[\Ag][\OO]-k_\text{L}[\Ag]
    \label{long_form}
\end{equation}
where $k_\text{L}$ is the non-reactive trap loss, and we assume that the apparent rate coefficient $k^\ast$ captures $k_r^\text{eff}$ and $k_3^\text{eff}$ via:
\begin{equation}\label{short_form}
    k^* = k_r^\text{eff} + k_3^\text{eff}[\ce{He}].
\end{equation}

\begin{figure}[h]
    \centering
    \includegraphics[width=3.33in]{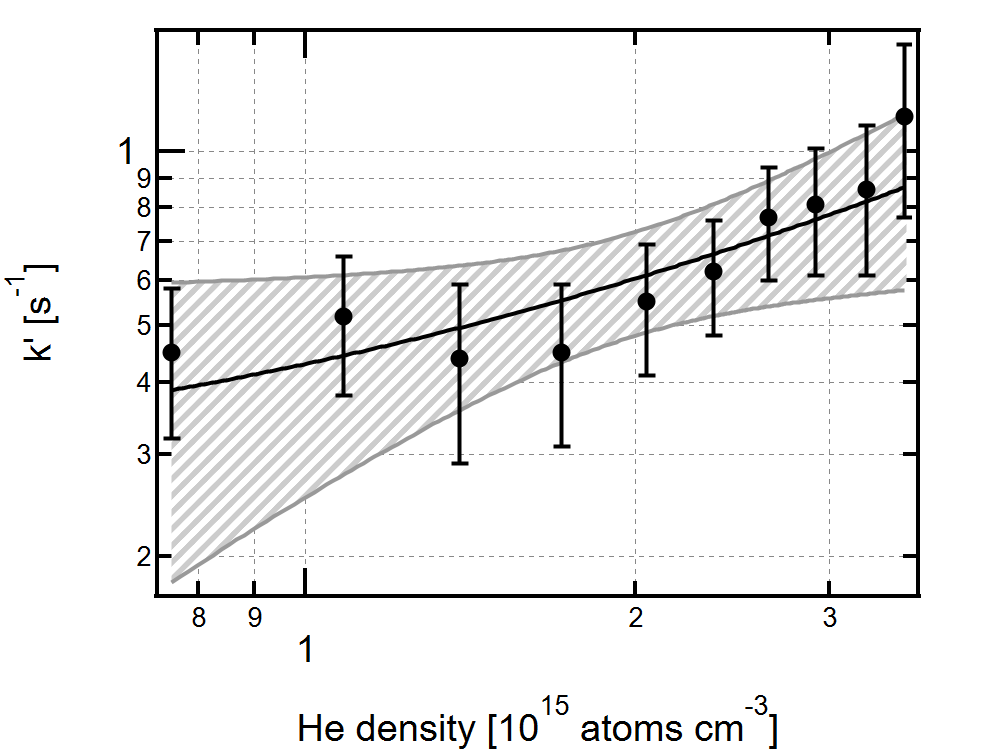}
    \caption{Dependence of the pseudo-first-order rate coefficient on helium buffer gas density in the trap. The fit yields the values of the effective radiative-association rate coefficient and the effective termolecular association rate coefficient. The shaded area shows the 95\% confidence interval band.}
    \label{fig:pressure}
\end{figure}

Fig.~\ref{fig:pressure} shows the apparent rate coefficient $k^\ast$  measured as a function of ion-trap pressure, from 15 to 110 mTorr (5\E{14} to 3.5\E{15} He atoms cm$^{-3}$). The pressure of the ion trap was varied by changing the flow of He while maintaining a constant oxygen concentration of $\sim 3$\E{12} \mcc. At the lowest pressures $k^\ast$ is constant with pressure, but starts to increase linearly above trap densities around 1\E{15} He atoms cm$^{-3}$. This trend is consistent with the onset of collisional stabilization via \eqref{rxn:Ter} with the lower-pressure measurements reflecting radiative stabilization. We fit the experimental data shown in Fig.~\ref{fig:pressure} using the function \eqref{short_form} to obtain the lower limit of $k_r^\text{eff} = 1$\E{-15} \rateUnits{} imposed by the trap stability and $k_3^\text{eff} = (5.8 \pm 5.1)$\E{-29} \terUnits{}. The data and fit are plotted along with the 95\% confidence bands to demonstrate the sensitivity of this model to our data. 

Koyanagi et al. measured the room-temperature rate coefficient for \ce{Ag+ + O2} using a selected-ion flow tube at 295 K and 0.35 Torr He buffer gas and reported an absolute uncertainty of approximately $\pm30\%$ on their value of $k^\ast = 1$\E{-13} \rateUnits under collisional conditions. The experiments were performed under pseudo–first-order conditions with \OO{} in excess; however, the absolute \OO{} number density was not reported in their study. The authors further note that slow \OO{} addition reactions are presumed to proceed via collisional stabilization under their conditions and that pressure-dependent studies were not performed. Given the very low reaction efficiency ($k^\ast/k_\text{capture} \sim 10^{-4}$ reported for \ce{Ag+ + O2} addition) and the sensitivity of such systems to collisional environment, internal energy distributions, and detection definition (parent loss vs stabilized product), factor-of-few differences between effective rate coefficients measured in different reactor geometries and pressure regimes are plausible. The present work makes a significant step forward by reporting the pressure dependence of this reaction.
 
\section{Conclusions and  Perspectives} \label{sec:conclusions}

We developed a new instrument that is tailor-made to directly measure rate coefficients and identify products from radiative-association reactions. Two critical components of GDIT are: (1) the bright and continuous glow-discharge source, that will be capable of producing a wide variety of atomic and molecular ions for study of ion--molecule reactions, and (2) the linear ion trap capable of trapping reactant and product ions for the long times needed to probe radiative-association processes. The collisional environments in the source and trap ensure thermalization of all reactant degrees of freedom. 

Radiative-association reactions have presented historical challenges to measuring reaction rate coefficients. The low rate coefficients typical of radiative association mean three-body collisional processes tend to dominate at typical laboratory conditions. Ion traps solve the inherent problems of slow reactivity. We demonstrate in this work the measurement of an effective radiative-association rate coefficient for the reaction between \ce{Ag+} and \ce{O2} with a lower limit of $1 \times 10^{-15}$ \rateUnits. This likely reflects the slowest effective rate coefficient we are currently equipped to measure, limited by our current trap stability. The direct detection of formed products will be critical in establishing mechanistic information in future studies. The next advance for GDIT will be to improve the mass resolution in product detection by incorporating a non-linear mass resolution function\cite{pedder_practical_nodate} in the kinetics data acquisition program to achieve $\Delta m/m$ values closer to 100, a necessary step for more complex reaction systems. If needed, a time-of-flight mass spectrometer could be added for more-precise product detection. 

The modularity of GDIT presents promising directions for future development. The present experiments were carried out at room temperature; future implementation of a cryogenic ion trap will enable the study of radiative-association processes at temperatures relevant to the astrophysical objects where they dominate. The temperature dependence seen in radiative association (via sensitivity of the rate to the rotational state of the reactant(s)) motivates the need for laboratory measurements at those temperatures, and provides important benchmarks for advanced theoretical efforts. These studies can also be extended beyond atomic ions, as molecular ions can also be readily produced in the glow-discharge ion source. The GDIT technique is therefore an adaptable means of studying radiative-association and other slow reactions, validating astrochemical models and producing new insight into the fate of charged species in astrophysical environments. 

\begin{acknowledgement}

This material is based upon work supported by the National Science Foundation under Grant No. 2154055. The authors thank Nicholas S. Shuman for help with LabVIEW programming, William S. Taylor for glow-discharge discussions, and Luke Metzler from Ardara Technologies for his assistance in instrument troubleshooting.

\end{acknowledgement}





\end{document}